# Lab-in-a-phone: Smartphone-based Portable Fluorometer for pH Field Measurements of Environmental Water

Md. Arafat Hossain, *Student Member, IEEE,* John Canning, *Member, IEEE,* Sandra Ast, Peter J. Rutledge, Teh Li Yen and Abbas Jamalipour, *Fellow, IEEE*

*Abstract*—A novel portable fluorometer combining the attributes of a smartphone with an easy-fit, simple and compact sample chamber fabricated using 3D printing has been developed for pH measurements of environmental water in the field. A colour filter attached over the camera white light LED selects an excitation band centred around $\lambda \sim 450$ nm with a 3 dB bandwidth, $\Delta\lambda \sim 21$ nm. An application-specific, temperature stable chemosensor based on the 4-aminonaphthalimide fluorophore was synthesized to absorb at this wavelength whilst emitting in the green region of visible spectra. The green emission is readily detected using the smartphone camera and a simple RGB Android application. Suppression of the green emission increases with increasing pH enabling a straightforward pH sensor. The system was calibrated against a commercial spectrofluorometer and pH measurements were taken at various locations around Sydney. The results were then compared directly with those obtained using conventional electrode based measurements. The data can be stored in the phone's available memory or transmitted by phone back to base for further real-time analysis.

## I. Introduction

THE measurement of pH in environmental water is of great importance because it is a facile indicator for the well-being of a system. Many organisms, including humans, cannot tolerate significant pH changes and acceptable ranges are defined by many bodies, including the World Health Organization (WHO) [1]. In practice, accurate pH measurement in water is accomplished with a sophisticated glass electrode connected to an electronic meter that needs to be repeatedly calibrated with standard buffers. This compromises their ease of use especially in resource-limited settings. Although microprocessor technology has made pH meters reasonably portable, they remain relatively specialised devices [2]-[4]. A simple low-cost and portable compact instrument would offer wider access and equitable distribution across even the poorest nations. A system that can be integrated into smartgrid sensor networks would offer further advantages, enabling direct access to central laboratory analysis (when combined with global internet accessibility). Here, we propose lab-in-a-phone sensing systems and demonstrate an example of one that meets all these concerns.

In recent years, smartphones have attracted significant attention as field-portable platforms for remote sensing. Many of the approaches reported exploit the inbuilt sensors such as gyroscopes and light metering within the phone. More recently, considerable attention has been focused on portable microscopies that take advantages of simple lens designs for imaging [5]-[11]. When combined with powerful software, phase information can be extracted (as with holographic processing) and ultrahigh spatial resolution (~ 100 nm) over extraordinarily far fields has been reported [12]-[14]. External high power diode sources were subsequently used to demonstrate fluorescence microscopy on a phone obtaining similarly unprecedented resolution over far field dimensions – these applications, however, all concentrate on using the smartphone as an imaging device [5]-[16]. These remarkable imaging systems with such large far fields are able to generate huge amounts of data accelerating the onset of a new frontier in information gathering, transmission, processing and storage that will feed on growing developments in fast communications and smartgrids. In terms of functionality beyond imaging, we have suggested, and demonstrated, that the evolution of high end organic light emitting diodes (OLEDs and AMOLEDs) now offers spectrally pure sources for potential fluorescence applications, including sensing [17]-[18]. It is clear that, the key drivers involve tremendous advances in compact wireless technology, including brighter diodes, white light sources and a variety of sensors such as more sensitive complementary metal-oxide semiconductor (CMOS) chips, global positioning system (GPS) modules, and so on, all of which are set to continue for some time. In addition to the portability advantages of a smartphone system, the open source nature of the Android platform means that customised program applications can be written to link

J. Canning, M. A. Hossain, S. Ast, P. J. Rutledge and T. L. Yen are with *i*nterdisciplinary Photonic Laboratories & School of Chemistry, The University of Sydney, NSW 2006, Australia. (e-mail: john.canning@sydney.edu.au; sandra.ast@sydney.edu.au; peter.rutledge@sydney.edu.au; t.liyen@yahoo.com).

A. Jamalipour and M. A. Hossain are with the Wireless Network Group, School of Electrical and Information Engineering, The University of Sydney, NSW 2006, Australia (e-mail: abbas.jamalipour@sydney.edu.au; mdarafat.hossain@sydney.edu.au).

together all the smartphone technology modules, heralding the potential development of advanced, low cost "lab-in-a-phone". The low cost is driven by the large telecommunications market (more than 7 billion mobile subscriptions by the end of 2013 signifying rapid global penetration rate of the smartphone [19]-[20]). More notably, this market is accessible and sufficiently affordable to extend into regions of the world that are traditionally technology deficient. Therefore, there is tremendous scope for combined lab-in-a-phone technologies for bio-analytical applications in remote and resource-poor settings, generating real-time results that can be remotely accessed through emerging smartgrid networks by, for example, analysts and certified professionals located in another continent. Needless to say such technological advances are extremely attractive for a number of fields, including environmental sensing and remote monitoring of infrastructure where field deployment is necessary.

In this work, a mobile fluorometer based on a customised smartphone diagnostic platform and an application-specific fluorescent chemosensor dye designed to measure pH in environmental water are described. Almost all of the smartphone based imaging platforms reported to date use external optical sources for imaging the samples because the smartphone source irradiance is low [5]-[16]. Although the active matrix OLED (AMOLED) display has previously been utilised as an in-built source for fluorescence sensing measurements [17]-[18], the overall irradiance is still too low at present raising demands on detection sensitivity. Furthermore, present technology does not permit split screen or multi-panel displays though this will likely be addressed in next generation smartphone systems. This limitation means it is not presently practical to maintain an optical source in one half of the screen whilst using the other as the display. There is also a higher degree of mechanical engineering involved because the camera and LED source are often away from each other or on opposite sides of the smartphone. On the other hand, the white LED flash of the CMOS camera has substantially greater irradiance and is separated from the imaging processing of the smartphone. It is also collocated with the camera, designed for optimal illumination during imaging.

The opto-mechanical hardware to support the sample cell for a sensor system is fabricated using 3D printing. An appropriate Android image analysis application was used to digitally process the fluorescent image of the application specific dye solution that is sensitive to pH. To demonstrate its effectiveness, we have conducted a number of measurements of water sites around Sydney. Other variants of this concept are clearly possible.

## II. METHOD

### A. Opto-mechanical Hardware Overview

The portable fluorometer was implemented on an Android-driven smartphone (Agora HD Smartphone from "Kogan", 1.2 GHz quad-core processor, Jelly Bean 4.2 OS and 8 MP Camera) [21]. Although it has similar spectrally pure

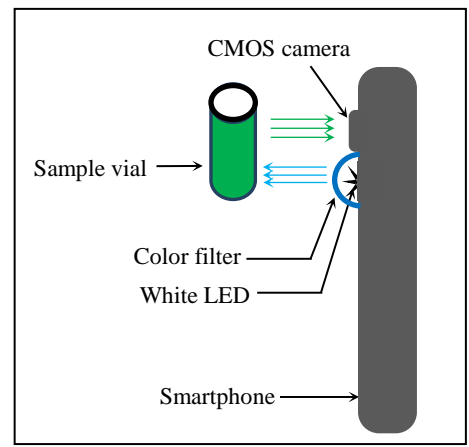

(a)

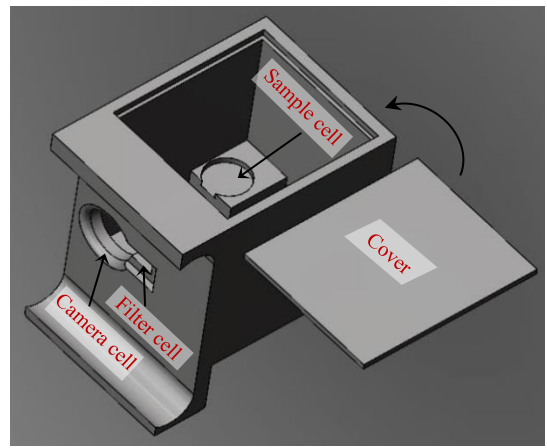

(b)

Fig. 1 (a) A schematic of the intended device configuration and; (b) the smartphone attachment designed in AutoCAD Inventor Fusion.

AMOLED emissions as our previously reported HTC phone platform [17], a much simpler alternative with more optical irradiance where required is the aforementioned white LED collocated with the CMOS camera. This avoids the need to use external sources in [5]-[16] as well as mirrors or other beam displacement hardware in [17]-[18]. The schematic summarising the principle of the smartphone sensor is shown in Fig. 1(a). In this fluorometer, the source is the white light LED normally used to illuminate the background of an image whilst the camera itself is the imaging sensor. The CMOS camera is already calibrated by the manufacturer to render as close as possible real-life colouring in images which are reproduced by the red-green-blue (RGB) AMOLEDs making up the display screen. For fluorometer operation, a blue thin film filter is placed over the white LED so that only blue light is transmitted to excite the sample placed in front of the camera. The emission band from the source is centered at 450 nm with a 3 dB bandwidth of ~21 nm, making it suitable as a fluorescent marker for the specific chemosensor dye.

3D printing offers the possibility that anyone, anywhere in the world can produce any object they need on demand [22-

24]. The 3D structure of the smartphone attachment is designed, in inventor software (AutoCAD Inventor Fission) shown in Fig. 1(b), to fit over the top of the camera unit. It was fabricated using a MakerBot Replicator 2X 3D printer. The attachment contains a sample cell chamber and suitable slots for the colour filter. To exclude the ambient light, the cell insertion side is closed by a suitable cover once the sample is in place. The whole assembly fixes firmly to the camera, is robust for transport, keeps the sample well shielded, and excludes light from external sources.

*B. Chemicals*

The mobile fluorometer relies on a chemosensor dye which changes fluoroscence intensity in response to the change in pH.

The application of a pH chemosensor has the advantage of rapid and reversible signal changes since protonation and deprotonation reactions occur faster than, for example, complex formation with metal ions. The probe was chosen to have an absorption and excitation wavelength ~ 440 nm and a large Stokes shift (emission in the green, at ~ 500 nm and higher). Chemosensors based on the 4-aminonaphthalimide fluorophore have relatively high quantum yields and high thermal- and photo-stability, reproducibly producing large signal changes and bright emission. Probes of this type (Fig. 2(a)) have been applied extensively in sensing applications due to their bright and robust spectral properties [25] and the ease of synthesis, enabling for the detection of a wide variety of analytes [26]-[30]. The photoindued electron transfer (PET) type 4-aminonaphthalimide dye 2-butyl-6-((2-(dimethylamino)ethyl)-amino)-1*H*-benzo[*de*]isoquinoline 1,3(2*H*)-dione, meets the spectral requirements of the mobile fluorometer and was synthesized according to the reported procedure [25]. The high quantum yield of the fluorophore, is caused by a charge transfer (CT) from the 4-amino group to the naphthalimide, the electron acceptor [31].

The excitation spectrum of the dye is centred at 440 nm with a 3dB bandwidth of ~55 nm (green) overlapping with the excitation source (blue) as shown in Fig. 2(b). The green fluorescence of the probe is low in the deprotonated form at higher pH by virtue of a PET from the dimethylamino-group (proton-acceptor) to the 4-aminonaphthalimide fluorophore. At lower pH, protonation of the dye leads to an enhanced emission as a function of pH down to pH ~ 4 (Fig. 5 (a)). Dyes of this type can be engineered to work across all pH regions as required.

Another attractive feature of this probe is the thermal stability under typical operating conditions – in Australia, environmental temperatures may vary from 15 to 50 °C. The thermal robustness was confirmed experimentally at different pH values between pH ~ 4 to 11 where the variation in fluorescence response, measured on the spectrofluorometer, was within experimental error over a temperature range of 22 to 40 °C shown in Fig. 3.

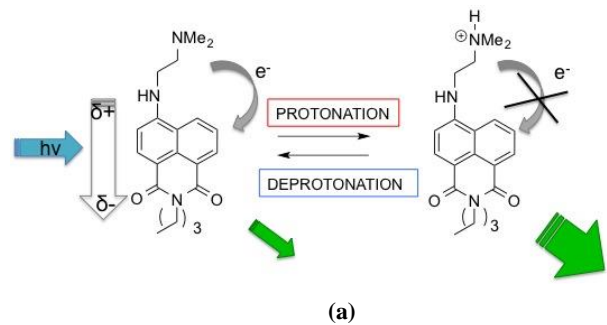

(a)

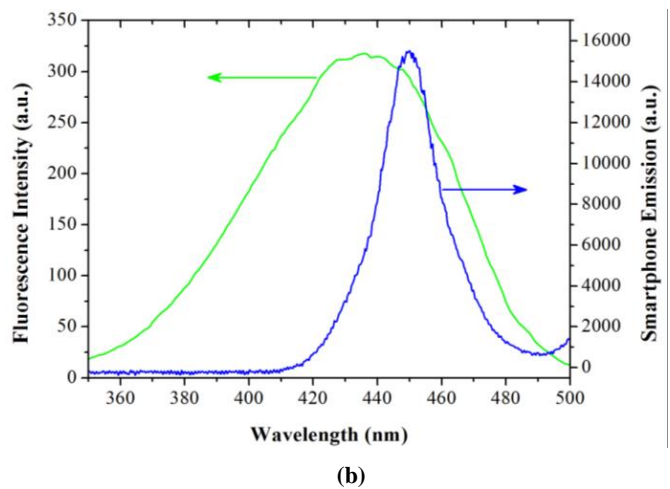

(b)

Fig. 2 (a) Fluorescence switching mechanism in PET-type 4-aminonaphthalimid-based pH probe; and (b) comparison of the probe, excitation (green line) with the mobile source emission (blue line).

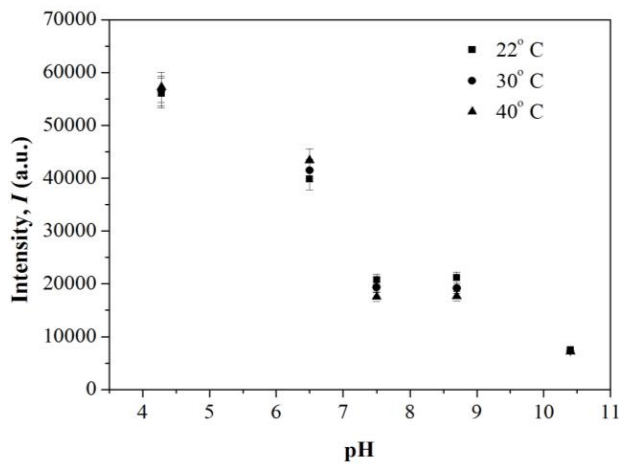

Fig. 3 Fluorescence intensity of the chemosensor with varying pH at three distinct temperatures. Within experimental error there is no significant change.

*C. Android-based Application*

To analyse the colour image taken by the CMOS camera, an image colour analysing app "Color Grab" was used [32]. It displays the RGB content of a live image in video mode in one corner of the image. As a test sample of the dye in water is excited, the value for green fluorescence intensity is readily detected within a fixed region of interest (see Fig. 4). The black sample cell holder removed external light and reduced any scattering of blue excitation light. This cell was further designed to position the sample vial in the centre of the white LED whilst imaged readily by the camera.

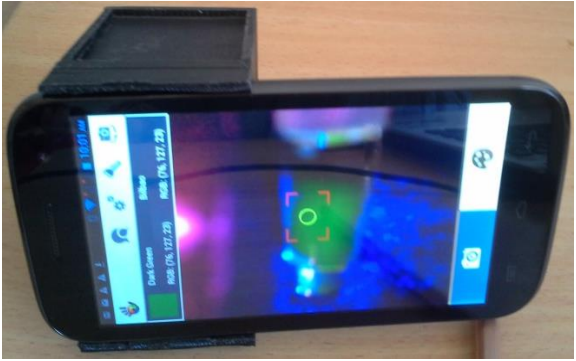

Fig. 4 Photograph of smartphone based fluorometer in action; a fluorescent green image is displayed on screen and the green intensity within a circular region of interest is directly displayed on top both corners of vertical screen.

A number of solutions across a wide pH range were prepared and the intensity of green emission was monitored with the mobile app. The trend analysis in these data gave very similar responses as compared to commercial spectrofluorometer measurements (Fig. 5(a)) and validating the mobile device for such pH-measurements. Data recorded on the smartphone fluorometer can be stored on phone memory or transmitted wirelessly for further analysis.

### III. SYSTEM CALIBRATION

In order to measure an unknown pH, the smartphone software must first be calibrated against reference samples. A total of 14 standard buffer solutions ranging from pH 4.28 to 10.40 were prepared using HCl and NaOH at an average temperature of 23 °C, and the pH of each confirmed using a standard pH electrode. The chemosensor was applied as a 2 mM solution in dimethyl sulfoxide (DMSO). Each sample solution was prepared by adding 8.0 μL of the 2 mM dye-solution to 1.59 mL of buffer (giving a final solution at ~ 10 μM dye) in glass vial of 1.2 cm diameter. Finally, the green fluorescence intensity of each sample was measured with the spectrofluorometer as well as the smartphone fluorometer. The smartphone app software was used to analyze each image taken. During each measurement, the vial was located at a distance of exactly 3 cm to the flash LED (Fig. 1) – this was kept constant for all measurements. Fig. 5(a) compares the fluorescence intensity measured using the smartphone with that made using the spectrofluorometer. The smartphone results are in good agreement with the spectrofluorometer and deviations are within experimental error. There is a linear trend of decreasing fluorescence intensity from pH ~7.0 to 9.7 in both sets of measurements. This defines the working range of the smartphone instrument. Fig. 5(b) plots the pH values with changing fluorescence intensity; from the linear fit of these data, the empirical equation relating pH to intensity is

$$pH = 10.766 - 0.02109 * I \quad (1)$$

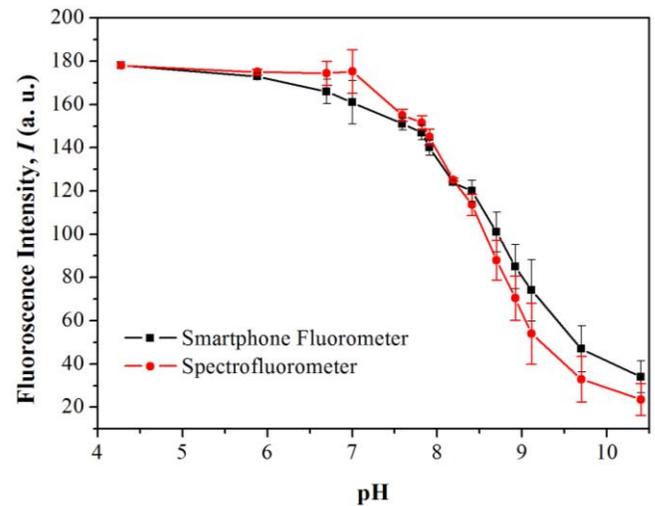

(a)

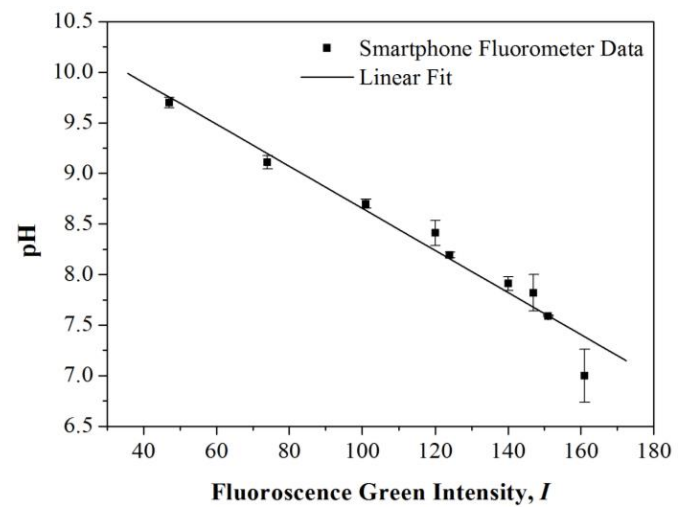

(b)

Fig. 5 Calibration of the mobile fluorometer against the commercial spectrofluorometer – (a) variation of fluorescence intensity with pH and; (b) linear fit of pH against fluorescence intensity on the mobile fluorometer.

This linear fit equation can then be used to quantify the target pH level in a given water sample between pH ~ 7 and 10 by measuring the fluorescence intensity created by the probe. pH measurements were then made using the smartphone fluorometer and (1) as well as a standard pH meter operating in the laboratory at an average temperature of 23 °C.

### IV. RESULTS AND DISCUSSION

The evaluation of the pH measuring platform was performed using three different types of sample water. A total of 11 environmental samples (7 × sea and 4 × lake water) were collected around Sydney (Fig. 6). Measurements with the smartphone fluorometer were made real-time at the site at which samples are collected. In order to check the accuracy of the results, the same samples were also measured back in the laboratory with a commercial, non-portable, high accuracy pH meter (METER TOLEDO, ±0.01). To preserve the integrity of the collected samples in transit, these were stored in borosilicate bottles filled to the top and transported to the lab under ice in a container, which also cuts out background light, within 2-3 hours. The samples were then brought to the same temperature as the ones recorded during the 'live' field measurements with the smartphone fluorometer. To assess the potential effects of salts and other solutes on the fluorescence intensity of the dye, the sensor was first tested against salt solutions in the laboratory using samples containing NaCl (0.2 M) and $MgCl_2 \cdot 6H_2O$ (0.2 M) (the most abundant dissolved species present in seawater [31]), at pH 8.41. Beside these, a tap water sample in the laboratory with unknown chemical composition was also tested. The fluorescence intensities of these samples were measured with identical dye concentration and experimental set up. All sets of data obtained are presented in Fig. 7 and 8.

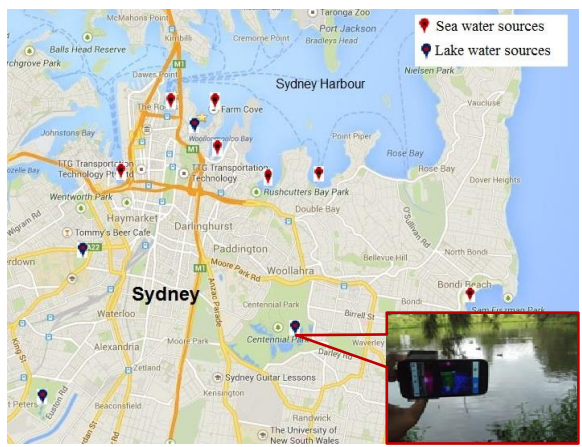

Fig. 6 Sample measurements were taken across Sydney as shown above.

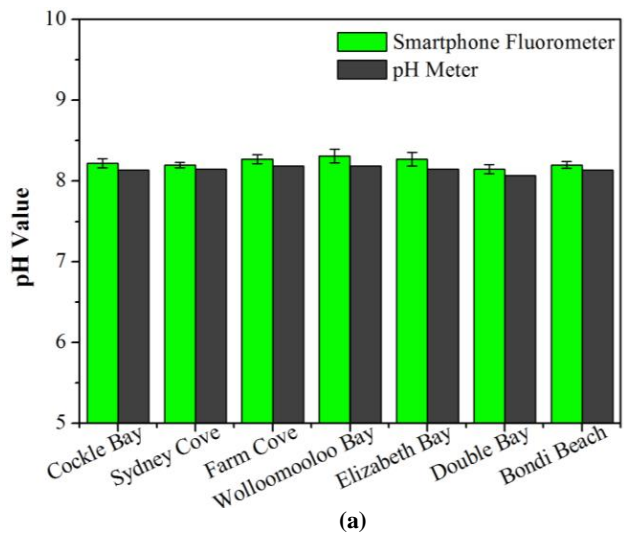

(a)

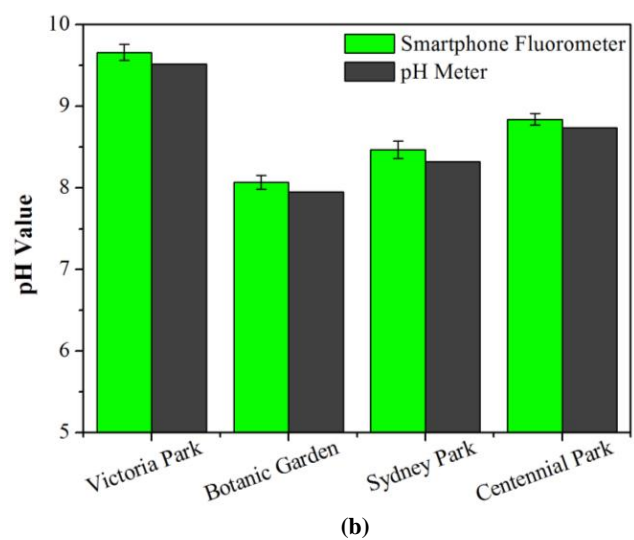

(b)

Fig. 8 Results of pH measurement for – (a) seawater and; (b) lake water. Measurements in both systems were performed at identical temperature for each sample.

Firstly, the dye was found to be sufficiently stable to work under various salts and chemicals commonly available in seawater. The results of such measurement are shown in Fig. 7. Comparing with standard pH meter measurements, it is clear that the dye used in the smartphone based detection platform is robust under various environmental conditions.

The pH values of seawater taken at different locations were measured on site with the smartphone fluorometer and the results are presented in Fig. 8(a). They are compared to standard electrode based measurements (made after transporting samples back to the lab). Using the mobile fluorometer the pH of the seawater samples were measured, giving results that are comparable with the electrode based system and in good agreement within a small experimental error. The water within these locations is found to be slightly basic, with pH ranging over (8.07 to 8.19) ± 0.12. This compares favourably to data reported using field portable electrode based pH meters [34]-[35].

Lake water samples were filtered with a common syringe filter to remove extraneous particulate materials such as fungi, a potential cause of background fluorescence. The pH measured in the lake waters are found to vary widely, from 44(7.95 to 9.52) ± 0.15, presumably a reflection of the different local ecologies.

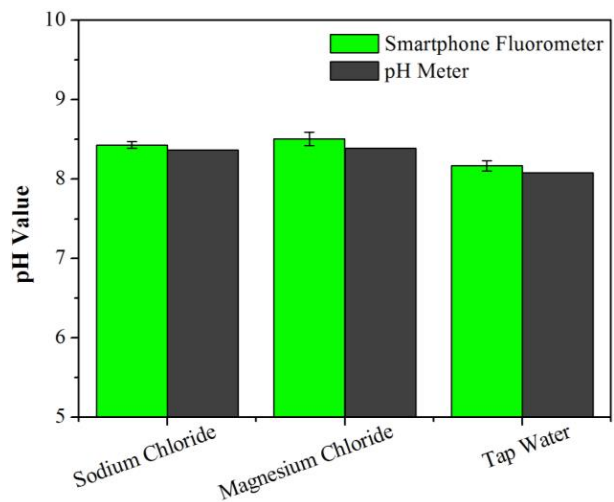

Fig. 7 Results of pH measurement for – lab water with different chemicals.

The measurement of pH can be extended to other areas by expanding the measurable range through the design of customised dyes for each environment. Finally, all collected data was sent wirelessly via email for archiving, demonstrating the potential of wireless data transmission and off-site analysis. This is ideal for sensors smartgrids collecting and emrging for analysis similar data from many portable instruments, something that is a cornerstone of the next generation internet of things. The demonstration reported here opens the way to quick, cost-effective spatio-temporal analysis of samples *in situ*, with instant reference to a database of statistical chemical properties elsewhere in the world. This approach could be especially useful for environmental water monitoring at remote sites on national and global scales.

## V. CONCLUSION

We have developed the first functioning field fluorometer using a smartphone platform. Other than the 3D printed sample holder, the source, interrogation and imaging unit are contained within the smartphone itself – no external sources are required. Through wireless communications, the mobile instrument can be networked into global smartgrids allowing instant access to additional analysis at a different location. The particular instrument was based on a pH-sensitive chemosensor that uses a robust fluorescent dye tailored to overlap with the blue spectral emission of typical smartphone OLEDs (or AMOLEDs), although greater irradiance was obtained using a filter and the white LED used for image illumination with the camera. This greatly relaxed the physical requirements of the sample cell, increased irradiance, and avoided any need for additional optical components such as mirrors. A mobile device like this also clearly eliminates the need for external hardware such as lasers and more sensitive cameras (although these could in theory be added) – this is key to broader technology accessibility and affordability.

This system successfully measured the pH of water from different sites in Sydney in good agreement with a conventional electrode-based measurement. The test results can either be stored in the smartphone memory or transmitted to a central computer for real time data processing. There is great potential to extend this approach to enable the detection of many species, including metal ions and other species of environmental concern. Furthermore, the technology has clear biological and biomedical potential and with greater focus on specific innovations can be significantly improved in terms of signal-to-noise detection. The convergence of these technologies makes the portable fluorometer the first true demonstration of a "lab-in-a-phone".

ACKNOWLEDGMENT

The authors acknowledge support from the Australian Research Council (ARC) through grants ARC FT110100116 and DP120104035. The Agora HD smartphone was provided by John Canning. Md. Arafat Hossain acknowledges an International Postgraduate Research Scholarship (IPRS) from the University of Sydney.

(7.95 to 9.52) ± 0.15, presumably a reflection of the different local ecologies.

The measurement of pH can be extended to other areas by expanding the measurable range through the design of customised dyes for each environment. Finally, all collected data was sent wirelessly via email for archiving, demonstrating the potential of wireless data transmission and off-site analysis. This is ideal for sensors smartgrids collecting and emrging for analysis similar data from many portable instruments, something that is a cornerstone of the next generation internet of things. The demonstration reported here opens the way to quick, cost-effective spatio-temporal analysis of samples *in situ*, with instant reference to a database of statistical chemical properties elsewhere in the world. This approach could be especially useful for environmental water monitoring at remote sites on national and global scales.

## V. CONCLUSION

We have developed the first functioning field fluorometer using a smartphone platform. Other than the 3D printed sample holder, the source, interrogation and imaging unit are contained within the smartphone itself – no external sources are required. Through wireless communications, the mobile instrument can be networked into global smartgrids allowing instant access to additional analysis at a different location. The particular instrument was based on a pH-sensitive chemosensor that uses a robust fluorescent dye tailored to overlap with the blue spectral emission of typical smartphone OLEDs (or AMOLEDs), although greater irradiance was obtained using a filter and the white LED used for image illumination with the camera. This greatly relaxed the physical requirements of the sample cell, increased irradiance, and avoided any need for additional optical components such as mirrors. A mobile device like this also clearly eliminates the need for external hardware such as lasers and more sensitive cameras (although these could in theory be added) – this is key to broader technology accessibility and affordability.

This system successfully measured the pH of water from different sites in Sydney in good agreement with a conventional electrode-based measurement. The test results can either be stored in the smartphone memory or transmitted to a central computer for real time data processing. There is great potential to extend this approach to enable the detection of many species, including metal ions and other species of environmental concern. Furthermore, the technology has clear biological and biomedical potential and with greater focus on specific innovations can be significantly improved in terms of signal-to-noise detection. The convergence of these technologies makes the portable fluorometer the first true demonstration of a "lab-in-a-phone".

ACKNOWLEDGMENT

The authors acknowledge support from the Australian Research Council (ARC) through grants ARC FT110100116 and DP120104035. The Agora HD smartphone was provided by John Canning. Md. Arafat Hossain acknowledges an International Postgraduate Research Scholarship (IPRS) from the University of Sydney.